\documentclass[%
 reprint,
 amsmath,amssymb,
 aps,
floatfix,
]{revtex4-2}

\usepackage{graphicx}
\usepackage{dcolumn}
\usepackage{bm}
\usepackage{upgreek}
\usepackage{xspace}

\newcommand{\dscorrdsBH}{\ensuremath{(d\sigma_\mathrm{corr}/dy)/(d\sigma_\mathrm{BH}/dy)}\xspace}
\newcommand{\dscorrdsBHtheta}{\ensuremath{(d^2\sigma_\mathrm{corr}/dy\,d\theta_\gamma)/(d^2\sigma_\mathrm{BH}/dy\,d\theta_\gamma)}\xspace}
\newcommand{\ydef}{\ensuremath{y=E_\gamma/E_e}\xspace}
\newcommand{\sx}{\ensuremath{\sigma_x}\xspace}
\newcommand{\sy}{\ensuremath{\sigma_y}\xspace}
\newcommand{\myparallel}{{\mkern3mu\vphantom{\perp}\vrule depth 0pt\mkern2mu\vrule depth 0pt\mkern3mu}}

\begin{document}

\preprint{APS/123-QED}

\title{When invariable cross-sections change: the Electron-Ion Collider case}

\author{Krzysztof Piotrzkowski}
 \email{krzysztof.piotrzkowski@cern.ch}
 \affiliation{Université catholique de Louvain, Centre for Cosmology,\\ Particle Physics and Phenomenology,\\ 1348 Louvain-la-Neuve, Belgium.}

\author{Mariusz Przybycien}%
 \email{mariusz.przybycien@agh.edu.pl}
\thanks{This work was partly supported by the AGH UST statutory task No. 11.11.220.01/4 within subsidy of the Ministry of Science and Higher Education.}
 \affiliation{%
AGH University of Science and Technology,\\ Faculty of Physics and Applied Computer
Science,\\ Al.~Mickiewicza 30, 30-059 Kraków, Poland.}%


\date{\today}

\begin{abstract}
In everyday research, it is tacitly assumed that the scattering cross-sections have fixed values for the given particle species, centre-of-mass energy, and particle polarizations. However, this assumption has been called into question after several observations of suppression of high-energy bremsstrahlung. This process will play a major role in experiments at the future Electron-Ion Collider, and we show here how variations of the bremsstrahlung cross-section can be profoundly studied there using the lateral beam displacements. In particular, we predict very strong increase of the observed cross-sections for large beam separations. We also discuss the relation of these elusive effects to other quantum phenomena occurring over macroscopic distances. In this context, spectacular and possibly useful properties of the coherent bremsstrahlung at the EIC are also evaluated.
\end{abstract}

\maketitle



A strong suppression of the bremsstrahlung process, $e^+e^-\to e^+e^-\gamma$, was observed in 1982 by the MD-1 experiment at the VEPP-4 collider in Novosibirsk, at the centre-of-mass energy of 3.6 GeV \cite{Blinov:1982vp}. Despite significant systematic errors, the evidence of increasing suppression with decreasing energy of bremsstrahlung photons was striking. Small lateral beam-sizes and large impact parameters were proposed as the origin of this unexpected phenomenon \cite{Tikhonov1982cand,Tikhonov1982}. In 1995, this MD effect was observed, at a five-sigma confidence level, at the first electron-proton collider HERA at DESY \cite{Piotrzkowski:1995xh}. There, however, the effect was much smaller, only of about 3\%, as in the ZEUS experiment only much higher photon energies were measured and, in addition, the lateral beam-sizes at HERA were much bigger than 10~$\upmu$m, the beam-sizes at VEPP-4. 

Despite these observations, the notion of effective cross-section variations due to finite transverse beam-sizes has been slowly gaining wide recognition. Somewhat paradoxically, in the case of bremsstrahlung, the beam-size effect increases with the energy of collisions, as yet bigger impact parameters start to play a role in this process. At the Large Electron-Positron (LEP) collider at CERN, it manifested itself in a spectacular way of a very beneficial extension of the beams’ lifetimes \cite{Burkhardt:1994wk}. For the non-colliding LEP beams, the ultimate limit of beam lifetime of about 60 hours was due to the Compton scattering of thermal photons off the beam electrons and positrons - the process measured directly at HERA \cite{Piotrzkowski:1995xh,Lomperski:1993aw}. However, in the case of colliding beams the beam lifetimes at LEP were ultimately limited by the electron-positron bremsstrahlung (or, in other words, by the radiative Bhabha scattering) – but, the observed beam lifetime of about 20 hours was much longer than the predicted 14 hours, based on the standard bremsstrahlung calculations. Nevertheless, to explain this bremsstrahlung suppression at LEP, an ad hoc cut-off mechanism due to large particle density was evoked~\cite{Burkhardt:1994wk}, ignoring the results from VEPP-4.

It seems that this lack of comprehension might stem from deeply rooted convictions, in particular from the definition of two-particle scattering cross-section, which boils down to this very simple relation used in physics over and over again:
\begin{equation}
R=L\sigma,
\label{eq:eq_1}
\end{equation}
where $R$ is the rate of events of a given binary process with the cross-section $\sigma$ (which can be obtained from the scattering amplitudes derived from the theory), and $L$ is the factor called luminosity which depends on intensities of fluxes of colliding particles. In practice, this formula is extremely useful since it factorizes the setup-dependent $L$ from the ``universal'' cross-section, which for the given particle species depends only on the center-of-mass energy and particle polarizations. The problem is that such a factorization is not always correct. One possibility of breaking it is very well known – if the density of a beam of particles (or of a ``target'') become too large, one cannot assume only binary interactions anymore. But there is that other possibility too, having nothing to do with the beam densities, but with a basic assumption used to arrive at such a factorization – assumption that the momenta of both colliding particles are ``completely fixed'', i.e. they can be described by pure and simple plane waves, corresponding to the beam-sizes infinitely large laterally. In reality, it is never so and any real beam has a finite transversal size, therefore, instead of the plane wave description, the proper wave packet formalism formally should be always used \cite{GoldbergerBook}.

\begin{figure}[t]
\includegraphics[width=0.28\textwidth]{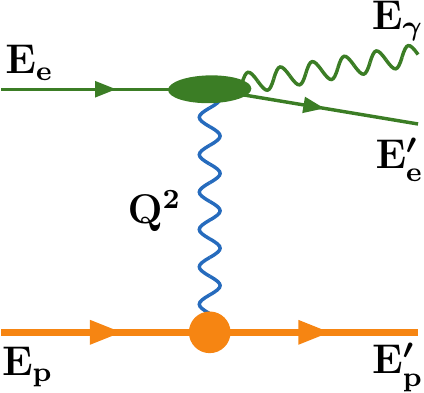}
\caption{\label{fig:fig_1} Feynman diagram for the electron-proton brems\-strahlung. Respective energy variables of incoming and outgoing particles are shown as well as the virtuality of exchanged photon $Q^2 = -(P_p - P_p^\prime)^2$, where $P_p$ and $P_p^\prime$ are four-momenta of the incoming and outgoing protons, respectively.}
\end{figure}

The bremsstrahlung process is so extraordinary due to extremely small momentum transfers between the radiating electron and the other charged particle, a proton in case of the electron-proton collisions, as in Fig. 1. In particular, it is kinematically allowed in this case that both incoming particles experience no angular scattering, that is, continue moving exactly along their initial directions, while the bremsstrahlung photon is emitted exactly in the direction of the electron momentum. It is precisely this configuration which provides the smallest virtuality $Q^2$ (see Fig.~\ref{fig:fig_1} for definitions of variables) of the exchanged photon:
\begin{equation}
|Q|_\mathrm{min}=m_e^2 m_p E_\gamma / (4E_p E_e E_e^\prime)
\label{eq:eq_2}
\end{equation}
where $m_e$ and $m_p$ are the electron and proton mass, respectively.

At high energies, this minimal photon virtuality acquires exceedingly small values – for example, at HERA $Q^2_\mathrm{min} = 10^{-8}$ eV$^2$ for $E_\gamma = 1$ GeV, and that results in a typical size of transverse momentum transfers, $q_\perp$, of about 0.0001 eV/$c$ only. This is because the bremsstrahlung differential cross-section, with unintegrated scattering angles, is proportional to $Q^{-4}$ (due to the propagator of exchanged photon), therefore the photon virtualities close to $Q^2_\mathrm{min}$ dominate, and to a very good approximation $Q^2 = Q^2_\mathrm{min} + q_\perp^2$. It should be stressed again that even $q_\perp = 0$ is allowed in bremsstrahlung. If one analyses this process in the impact parameter space \cite{Kotkin:1992bj}, these very small values of $q_\perp$ correspond to very large values of the impact parameter $b = \hbar/q_\perp$. That, in turn, explains why the original Bethe-Heitler cross-section calculations in the Born approximation are so precise, despite neglecting the size of the proton and its spin. In fact, the impact parameter in high-energy bremsstrahlung reaches really macroscopic values – the above example from HERA corresponds to $b$ of about 2~mm! And that precisely is the origin of the beam-size effect, or more generally of the observed modifications of bremsstrahlung cross-sections. In deriving Eq.~\eqref{eq:eq_1} it is assumed that both colliding beams can be properly represented by simple plane waves, which means that the assumed distribution of the impact parameter is uniform. However, when both colliding beams are strongly focused at the interaction point then that is not valid anymore as the large impact parameters are suppressed and the bremsstrahlung differential cross-section is ``over-sampled'' at the low impact parameters, where it is smaller. As a result, one observes an effective suppression of bremsstrahlung, increasing with decreasing photon energies, since the typical $q_\perp$ is proportional to $E_\gamma$, see Eq.~\eqref{eq:eq_2}.

\begin{figure}[t]
\includegraphics[width=0.48\textwidth]{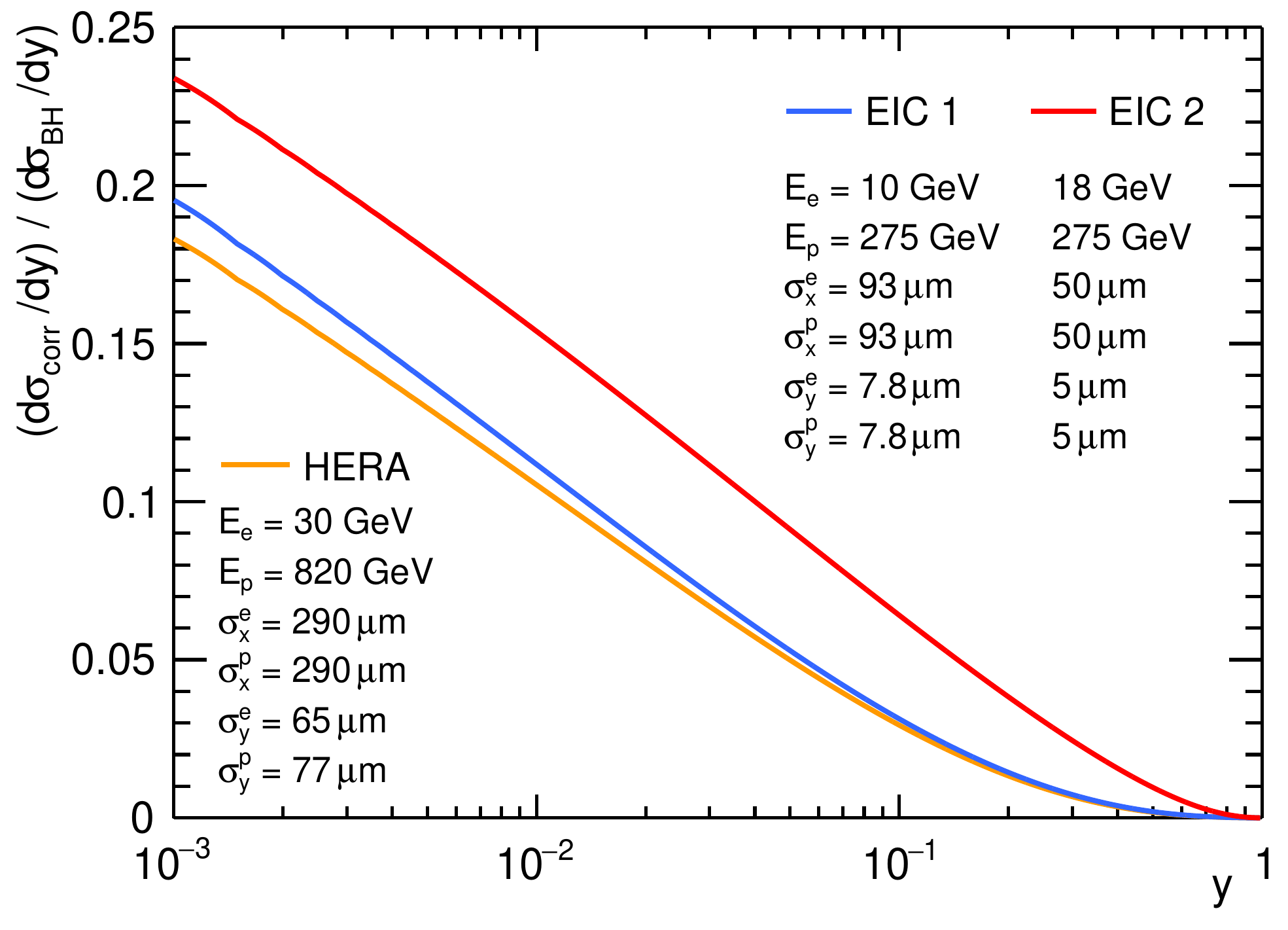}
\caption{\label{fig:fig_2} Relative corrections to the standard Bethe-Heitler cross-sections due to the beam-size effect. Relative suppression due to the beam-size effect \dscorrdsBH is shown as a function of \ydef for three cases of electron-proton bremsstrahlung. The corresponding beam energies and Gaussian lateral beam-sizes at the interaction point are listed. The curves were obtained using Eqs.~\eqref{eq:eq_dscorr_new} and \eqref{eq:eq_dsBH} from  Supplemental Material.}
\end{figure}

In Fig.~\ref{fig:fig_2} the relative correction \dscorrdsBH to the standard Bethe-Heitler cross-section, that is the bremsstrahlung suppression due to beam-size effect, is shown as a function of \ydef, where it is assumed that $d\sigma_\mathrm{obs}/dy = d\sigma_\mathrm{BH}/dy - d\sigma_\mathrm{corr}/dy$. Three cases are considered – one at HERA and two at the future Electron-Ion Collider (EIC). Despite significantly larger beam energies, the effect at HERA is smaller than these at the EIC. This is because of much stronger beam focusing at the EIC, in particular in the vertical direction – ``flat'' beams (with a large aspect ratio of at least 10:1) will be used there to avoid disruptive beam-beam effects, as the beam intensities will be much higher than those at HERA, what will result in the increase of luminosity by more than two orders of magnitude \cite{Willeke2019}. For the 18 GeV electron case, assuming extreme focusing, the bremsstrahlung suppression at the EIC will be of about 10\% for 1 GeV photons, and still of almost 1\% at $y = 0.5$.

Very precise cross-section measurements are essential for the success of the EIC scientific program. To achieve the requested 1\% precision, yet more accurate determination of the collider luminosity is necessary – Eq.~\eqref{eq:eq_1} can be used to measure cross-sections as well as to determine $L$ if a certain $\sigma$ is very well-known and the corresponding $R$ is accurately measured. As at HERA, the bremsstrahlung process will be used to precisely measure the EIC luminosity, therefore a thorough verification of the corrections due to the beam-size effect is mandatory. In principle, accurate measurements of bremsstrahlung angular distributions could provide interesting insights into the beam-size phenomenon [see Figs.~\ref{fig:fig_7} and \ref{fig:fig_8} in Supplemental Material]. That is however not possible to perform as the typical bremsstrahlung emission angles are much smaller than the angular divergences of beams at the EIC interaction point. Here, we propose a new, powerful and straightforward way of conducting at the EIC unique studies of the role of large impact parameters in bremsstrahlung, and of the origin of the beam-size effect in particular.

\begin{figure}[t]
\includegraphics[width=0.48\textwidth]{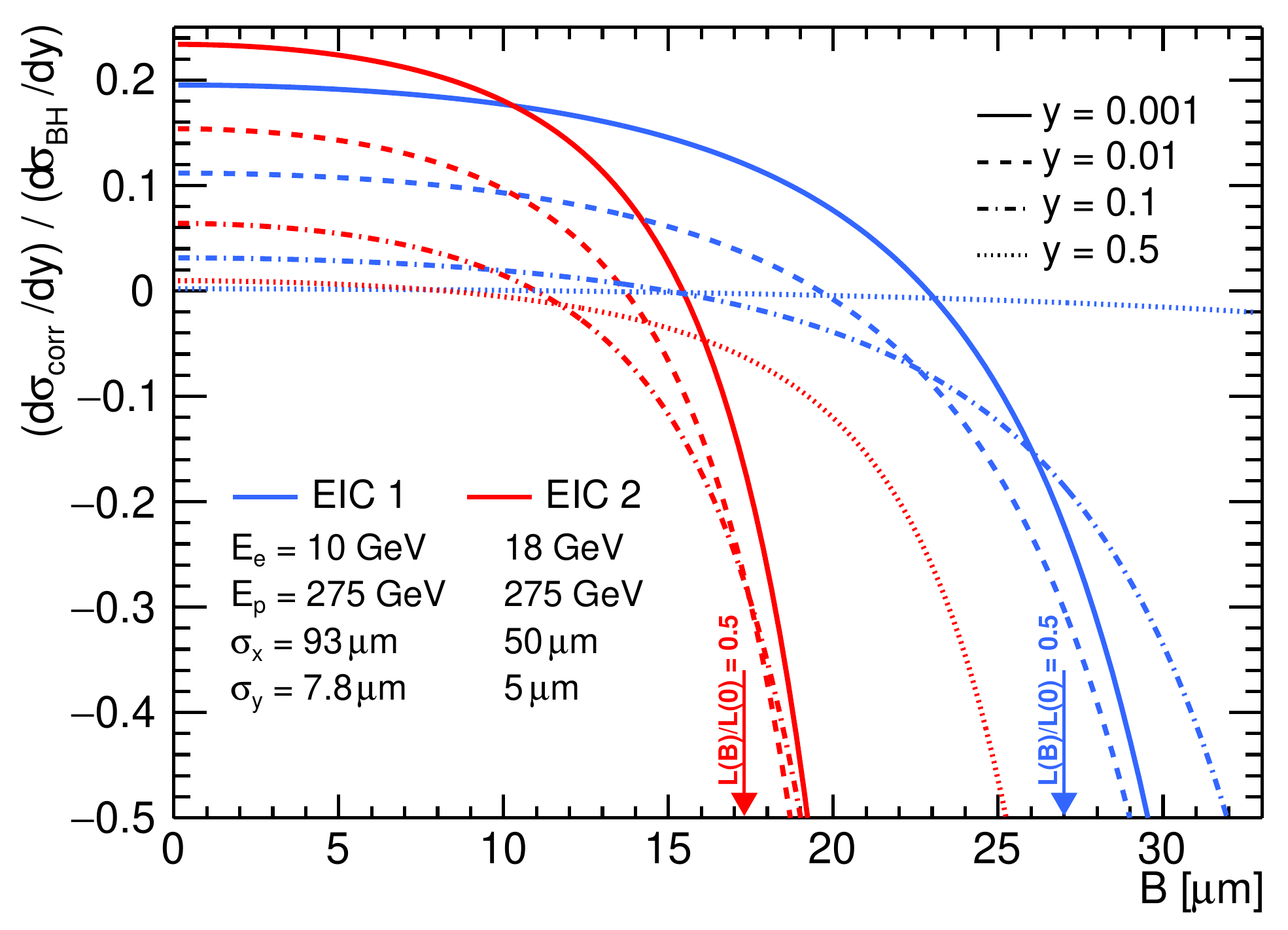}
\caption{\label{fig:fig_3} Relative corrections to the standard Bethe-Heitler cross-sections due to beam displacements at the EIC. The ratio \dscorrdsBH is shown as a function of vertical beam displacement $B$, for four values of $y$ and for two cases of electron-proton collisions at the EIC. Values of $B$ corresponding to the 5\% of the nominal luminosity are indicated and the beam energies and Gaussian lateral beam-sizes at the EIC interaction point are listed. The curves are obtained using Eqs.~\eqref{eq:eq_dscorr_new} and \eqref{eq:eq_dsBH} from  Supplemental Material with $G(\omega)$ given in Eq.~\eqref{eq:eq_3}.}
\end{figure}

We propose to precisely measure the bremsstrahlung spectra as a function of lateral beam displacements at the interaction point – that is, by using the so-called van der Meer scans, usually performed (at the LHC, for example) to measure the lateral sizes of the ``luminous region''. Such scans result in strong changes of the collider luminosity $L$ according to the formula, assuming Gaussian distributions of beam particles: $L(B) = L(0)	\exp{(-B/(\sigma_1^2+ \sigma_2^2)/2)}$, where $B$ is the lateral displacement of one of the beams within the horizontal or vertical plane, $\sigma_1$ and $\sigma_2$ are the two Gaussian widths in a given plane, often equal, and $L(0)$ corresponds to the luminosity of nominal, head-on collisions. However, in the case of bremsstrahlung, also its photon spectrum will be modified in a very specific way – the lateral beam displacement will serve as a direct scanner of impact parameter after all!

To make specific predictions, we extended the well-developed framework of Ref.~\cite{Kotkin:1992bj} - we also assume head-on beam collisions and Gaussian distributions of the beam particle coordinates in the transverse plane, $n(\vec{r}_\perp)$, but we allow for relative, lateral beam displacements. In particular, we introduced a vertical beam displacement $B$ of one of the colliding beams:
%
$$
n(\vec{r}_\perp) = \frac{N}{2\pi \sigma_x\sigma_y}\exp{\left(-\frac{x^2}{2\sigma_x^2}-\frac{(y-B)^2}{2\sigma_y^2}\right)}
$$
%
\noindent
where $\vec{r}_\perp =(x, y, 0)$, $N$ is the number of particles in a beam and \sx, \sy are the standard deviations in the horizontal and vertical planes, respectively. This resulted in the replacement of the original Eq. (5.32) by the following formula:
\begin{widetext}
\begin{equation}
G(\omega) = 2\int_{0}^{\infty}{\frac{\rho_\perp}{\rho_m} K_1^2\left(\frac{\rho_\perp}{\rho_m}\right)\left[1-\frac{e^{-v_+}}{\pi}\int_0^\pi{e^{v_-\cos{\varphi}}\cosh{\left(t_y\sin{\left(\frac{\varphi}{2}\right)}\right)d\varphi}}\right]}\frac{d\rho_\perp}{\rho_m}
\label{eq:eq_3}
\end{equation}
\end{widetext}
\noindent
where $G$ is the function which describes the number of missing equivalent photons of energy $\omega$ (see Eq. (5.14) in Ref.~\cite{Kotkin:1992bj}), $\rho_m = E_p/(m_p \omega)$, $t_y =\rho_\perp B/a_y^2$ and $v_\pm = \rho_\perp^2 (1 \pm a_y^2/a_x^2)/(4a_y^2)$, where $a_x^2 = \sigma_{x1}^2 + \sigma_{x2}^2$ and $a_y^2 = \sigma_{y1}^2 + \sigma_{y2}^2$ and $\sigma_{x1}$, $\sigma_{y1}$ and $\sigma_{x2}$, $\sigma_{y2}$ are, respectively, the horizontal and vertical beam sizes at the interaction point (that is, at $z = 0$) for two colliding beams; $K_1$ is the modified Bessel function of the second (third) kind.

For $B = 0$ the expression in the square brackets becomes: $[1 - e^{-v_+} I_0(v_-)]$, as in the Eq. (5.32) in Ref.~\cite{Kotkin:1992bj}, where $I_0$ is the modified Bessel function of the first kind.

\begin{figure}[t]
\includegraphics[width=0.48\textwidth]{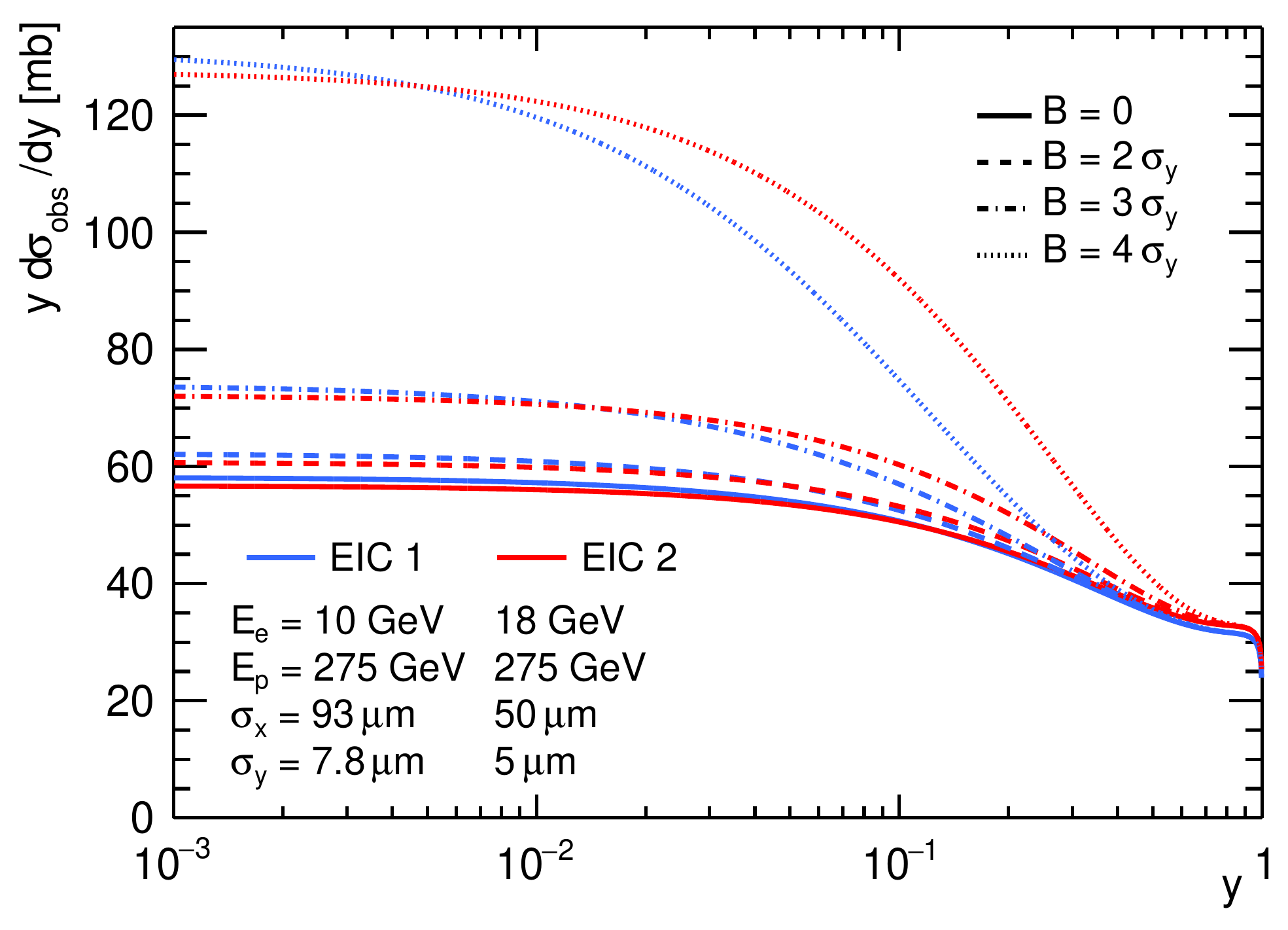}
\caption{\label{fig:fig_4} The predicted spectra of $ep$ bremsstrahlung at the EIC for several vertical beam displacements. The standard Bethe-Heitler cross-section $d\sigma_\mathrm{BH}/dy$ is modified due to the beam-size effect and beam displacements $B$. The effective cross-sections (multiplied by $y$ for better visibility) are shown for two cases of electron-proton collisions at the EIC - the corresponding beam energies and Gaussian lateral beam-sizes at the interaction point are listed. The vertical beam displacements $B = 2, 3$ and $4 \sigma_y$ correspond, respectively, to 36, 11 and 2\% of the nominal luminosity for $B = 0$. The curves are obtained using Eqs.~\eqref{eq:eq_dscorr_new} and \eqref{eq:eq_dsBH} from  Supplemental Material with $G(\omega)$ given in Eq.~\eqref{eq:eq_3}.}
\end{figure}

\begin{figure*}[t]
\includegraphics[width=0.48\textwidth]{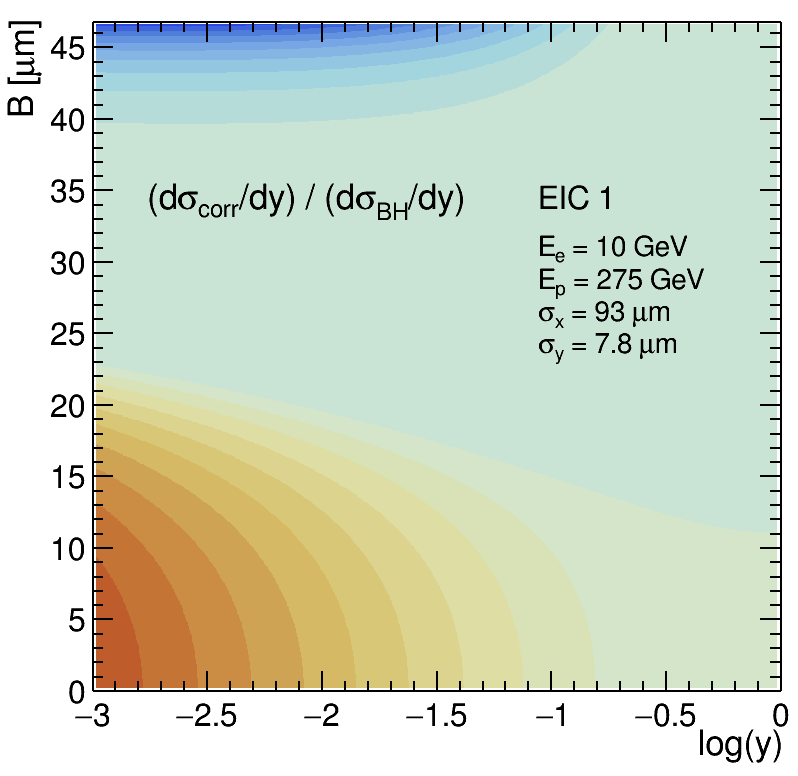}
\includegraphics[width=0.48\textwidth]{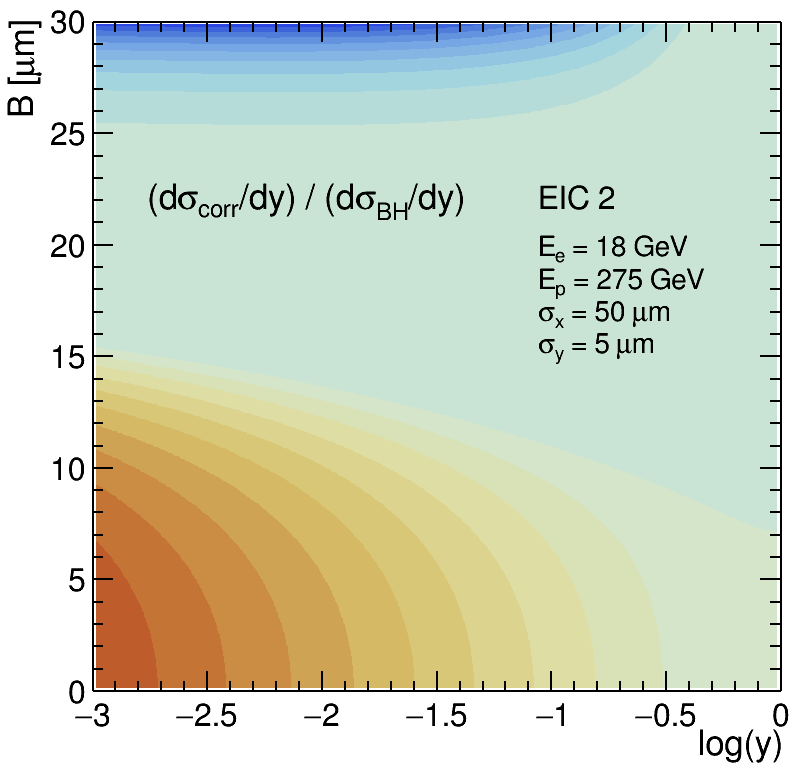}
\caption{\label{fig:fig_5} Relative corrections to the standard Bethe-Heitler cross-
sections, due to both the beam-size effect and vertical beam displacements, as a function of $B$ and $y$. The ratios \dscorrdsBH are shown as a function of the vertical beam displacement $B$ and the logarithm of the relative photon energy \ydef for the two sets of EIC parameters: EIC 1 and EIC 2 - the corresponding beam energies and Gaussian lateral beam-sizes at the interaction point are listed. Shown are 10 equidistant (in the third dimension) contours for the values above zero (displayed in brownish colours) and 10 equidistant contours for the values below zero (displayed in bluish colours). For the EIC 1 case, the distribution extends in the third dimension between approximately $-84$ and $+0.2$, whereas for the EIC 2 case this range is approximately from $-80.5$ to $+0.24$.}
\end{figure*}

In Fig.~\ref{fig:fig_3} we show how the relative cross-section corrections will change for four values of $y$, in the case of vertical beam scans at the EIC. As expected, for large values of $B$ the corrections become negative, which means that the bremsstrahlung cross-section will strongly increase – because the large impact parameters, for which the differential cross-section is increasing, are ``over-sampled'' in this case. In Fig.~\ref{fig:fig_4} we show how whole bremsstrahlung spectra are heavily distorted at the EIC for vertical beam displacements of only $20-30\,\upmu$m – in these cases the observed cross-section will double for low photon energies. For yet larger beam displacements the observed increase will become almost hundred-fold, see Fig.~\ref{fig:fig_5}, where we present the two-dimensional distributions of \dscorrdsBH as a function of both \ydef and the vertical beam displacement $B$. One should note however, that even for large values of $B$ the suppression due to the beam-size effect is still significant for small photon energies. Finally, from the experimental point of view, it seems preferable that such scans will be done at (very) low beam intensities to avoid beam disruptions, during the scans, due to the strong beam-beam effects.

A good understanding of the role of very large impact parameters in bremsstrahlung is mandatory for the precise luminosity determination at the future EIC, or for the proper explanation of the beam lifetimes at the KEK B-factory \cite{Kotkin:2005rm}, but it might also be relevant in non-accelerator physics. For example, it has been proposed that rapidly spinning neutron stars are the principal source of ultra-high energy cosmic rays. It is interesting to observe, if only for the illustration purpose, that the typical impact parameter for a 50~EeV electron radiating a 0.5~EeV photon reaches 40~km, which is about four times the neutron star radius. According to the results we obtained, radiation of such a photon can be either strongly suppressed or amplified, depending on the geometrical aspects of particle collisions.

Above, some new effects due to very small transverse momentum transfers are discussed, but there are also bremsstrahlung phenomena related to small longitudinal momentum transfers $q_\myparallel$, as the famous Landau-Pomeranchuk-Migdal effect \cite{Klein:1998du}, for example. In this case, these small transfers correspond to large coherence lengths (called often the formation length in this context) of the process, resulting in the bremsstrahlung suppression in dense materials if over such a longitudinal distance the incoming electrons undergo significant angular scattering \cite{Anthony:1995fs,Anthony:1995vv,Anthony:1997ed}. In high energy electron-proton collisions this coherence length $l_c$, calculated in the laboratory reference frame, is given by the formula: $l_c = \hbar/q_\myparallel = 4\hbar c\gamma_e^2/E_\gamma$, where $\gamma_e = E_e/m_e$. It was realized that if that coherence length is significantly larger than the length $\sigma_z$ of an opposite bunch, then the coherent bremsstrahlung takes place and beam electrons interact with the total field of all particles in this opposite bunch! The photon energy for which $l_c = \sigma_z$ is called the critical energy $E_c$.\\
\indent
Such a coherent effect results in a spectacular increase by many orders of magnitude of the bremsstrahlung cross-section for $E_\gamma \ll E_c$. For these low energies, the photon spectrum (per single bunch-crossing) can be simplified as follows \cite{Engel:1995pu}: $dN_\gamma = N_0~dE_\gamma /E_\gamma$, where $N_0 = 8\alpha N_e (r_e N_p /\sigma_x)^2/(9\sqrt{3})$ assuming flat beams with identical transverse beam-sizes, and $\alpha$ is the fine structure constant, $r_e$ is the electron classical radius, $N_e$ and $N_p$ are the electron and proton bunch populations, respectively, and $\sigma_x$ is the horizontal beam-size. As expected, the coherent bremsstrahlung per bunch collision scales with $N_e N_p^2$ but it has also a strong dependence on $\sigma_x$. At the EIC, due to high beam intensities and small beam-sizes, it will manifest itself in a unique and spectacular way. For example, one obtains $N_0 = 3.1(0.28) \times 10^{10}$ for these two sets of the EIC parameters: $E_e = 18(10)$ GeV, $N_e = 6.2(17.2) \times 10^{10}$, $N_p = 2.05(0.69) \times 10^{11}, \sigma_x = 50(93)\,\upmu$m and $\sigma_z = 6$~cm. The corresponding critical energy $E_c = 16.3$~keV and 5~keV, whereas the total radiated power in this process is about 900 and 100 W, respectively. In coherent bremsstrahlung the photon energies are relatively small but on average every single beam electron will emit more than one such an X-ray photon during every single bunch crossing. The radiation extends down to the ``visible widow'' - there, the coherent bremsstrahlung signal at the EIC will manifest itself as a flash of blue light lasting a small fraction of 1~ns, repeated up to 100 million times per second. For a human eye, at a spot where it hits the beam-pipe, its peak illuminance will reach 10 000 lux!

These results demonstrate that on the one hand unique studies of the coherent bremsstrahlung will be also possible at the EIC, as well as its potential applications for the beam diagnostics, but on the other hand, at the highest electron energy this coherent process will require a special treatment due to large radiated power, which in addition will almost double for significant vertical beam displacements \cite{Engel:1995pu}, such as those discussed above.
%
%
%
\bibliography{when_inv_cs_change}
\section*{Supplemental Material}
\begin{figure*}[t]
\includegraphics[width=0.48\textwidth]{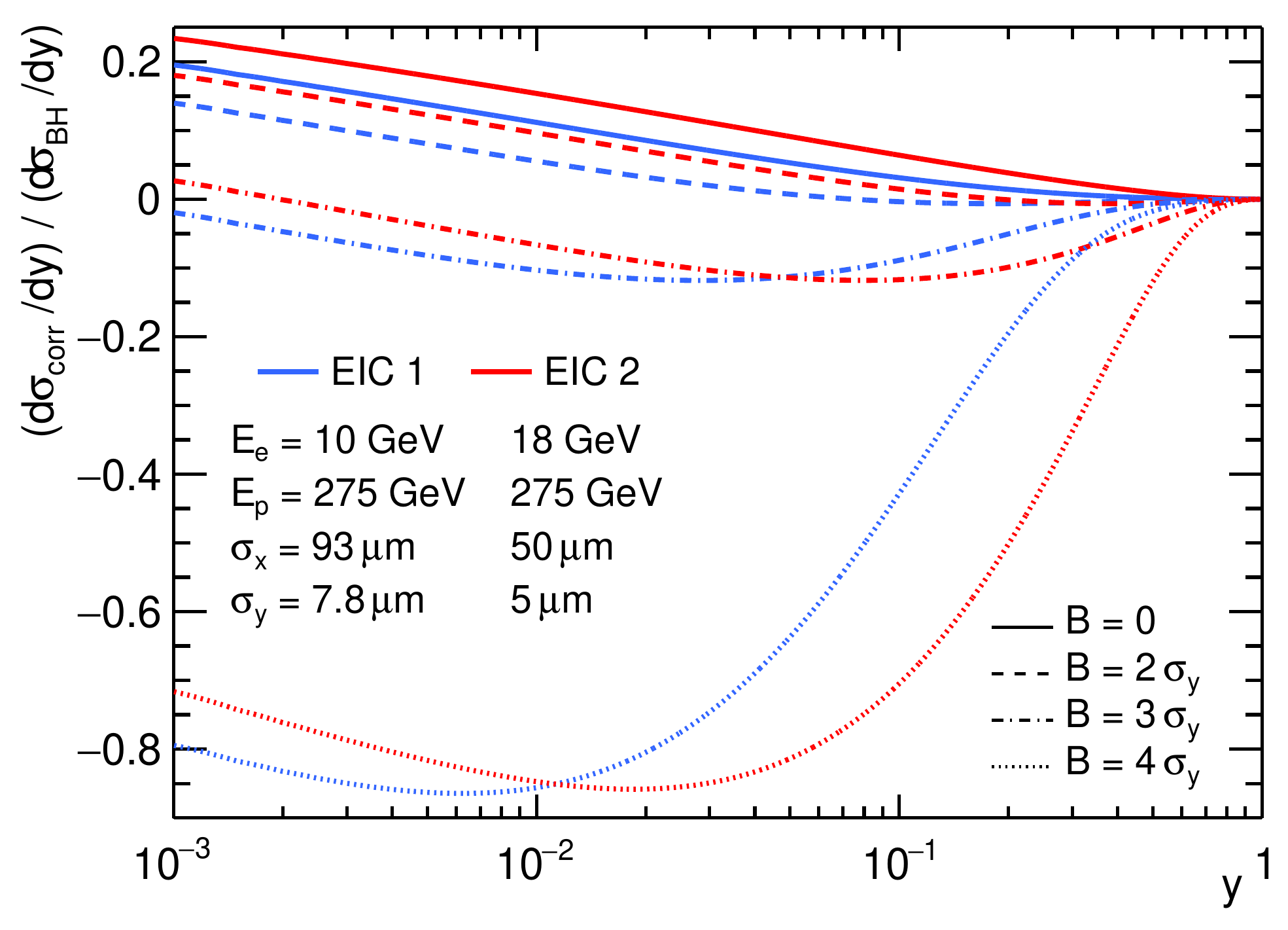}
\includegraphics[width=0.48\textwidth]{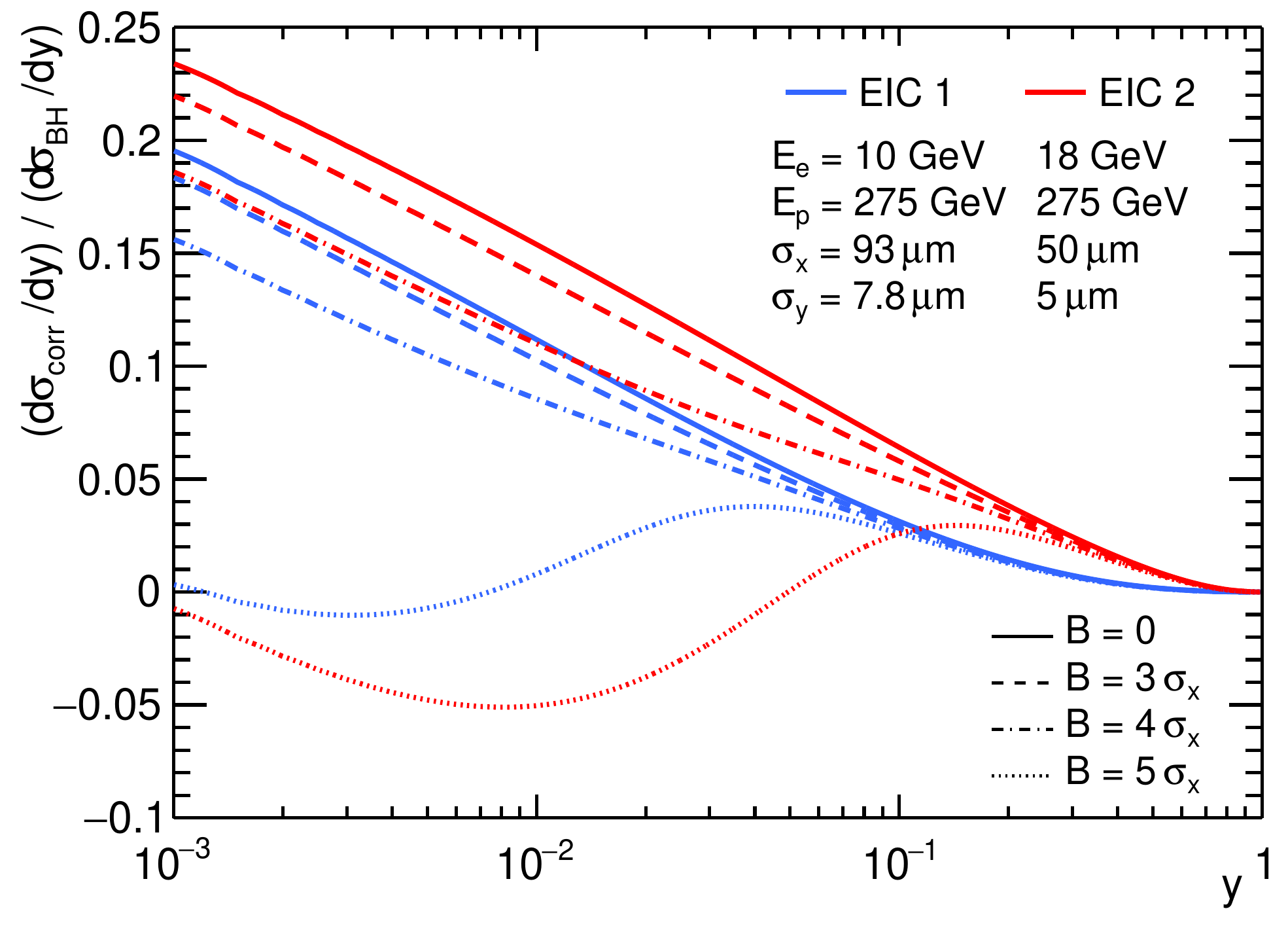}
\caption{\label{fig:fig_6} Relative corrections to the standard Bethe-Heitler cross-sections due to both the beam-size effect and (left) vertical or (right) horizonthal beam displacements. The ratios \dscorrdsBH are shown as a function of \ydef, for four vertical beam displacements $B = 0, 2\sigma_y, 3\sigma_y$ and $4\sigma_y$, and for four horizonthal beam displacements $B = 0, 3\sigma_x, 4\sigma_x$ and $5\sigma_x$. In both cases the two sets of EIC parameters are considered - EIC 1 and EIC 2 - the corresponding beam energies and Gaussian lateral beam-sizes at the interaction point are listed. The vertical beam displacements $B = 2, 3$ and $4\sigma_y$ correspond, respectively, to $36, 11$ and $2\%$ of the nominal luminosity for $B = 0$. The horizontal beam displacements $B = 3, 4$ and $5\sigma_x$ correspond, respectively, to $11, 2$ and $0.2\%$ of the nominal luminosity for $B = 0$.}
\end{figure*}

\begin{figure*}[t]
\includegraphics[width=0.48\textwidth]{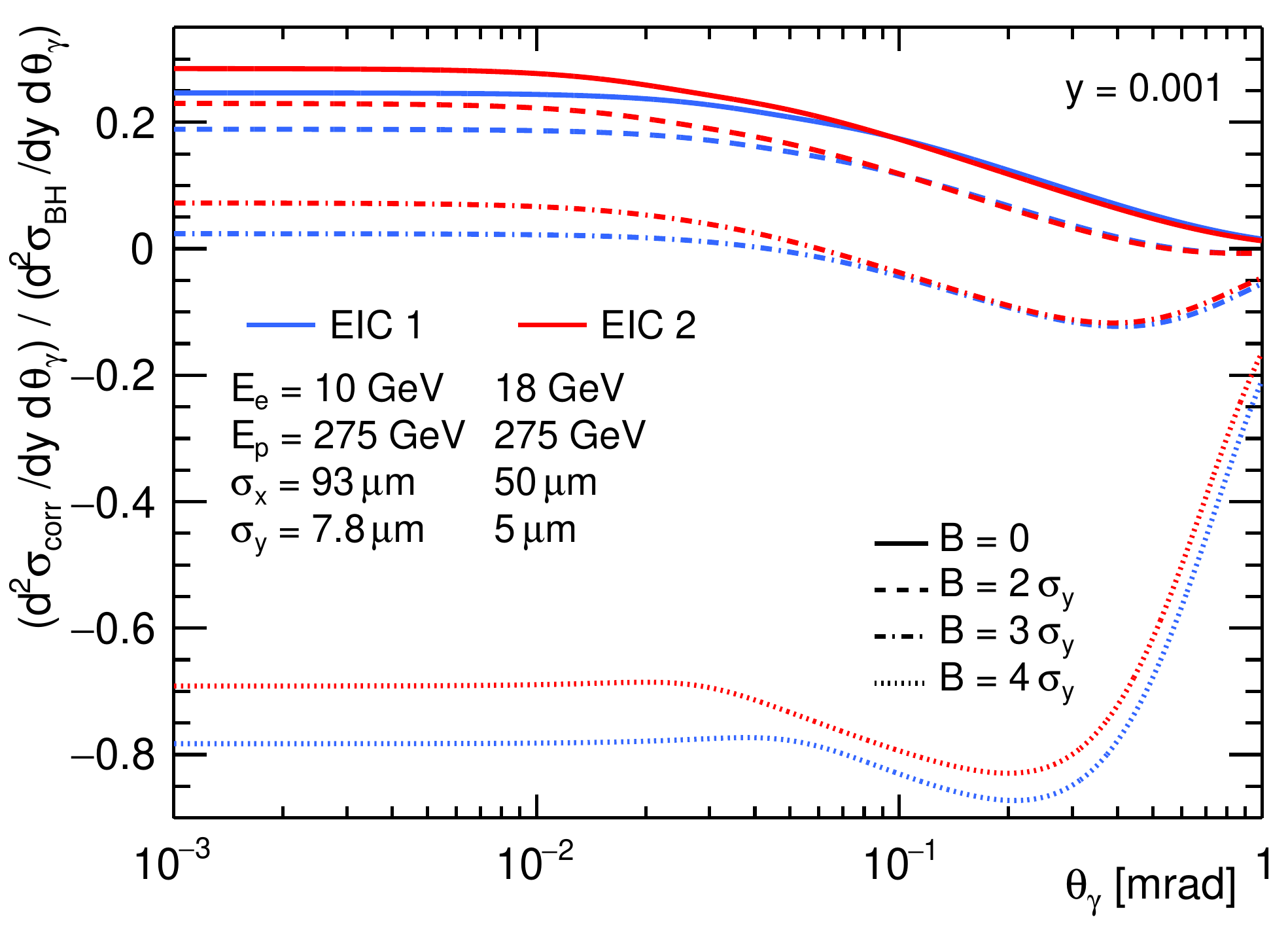}
\includegraphics[width=0.48\textwidth]{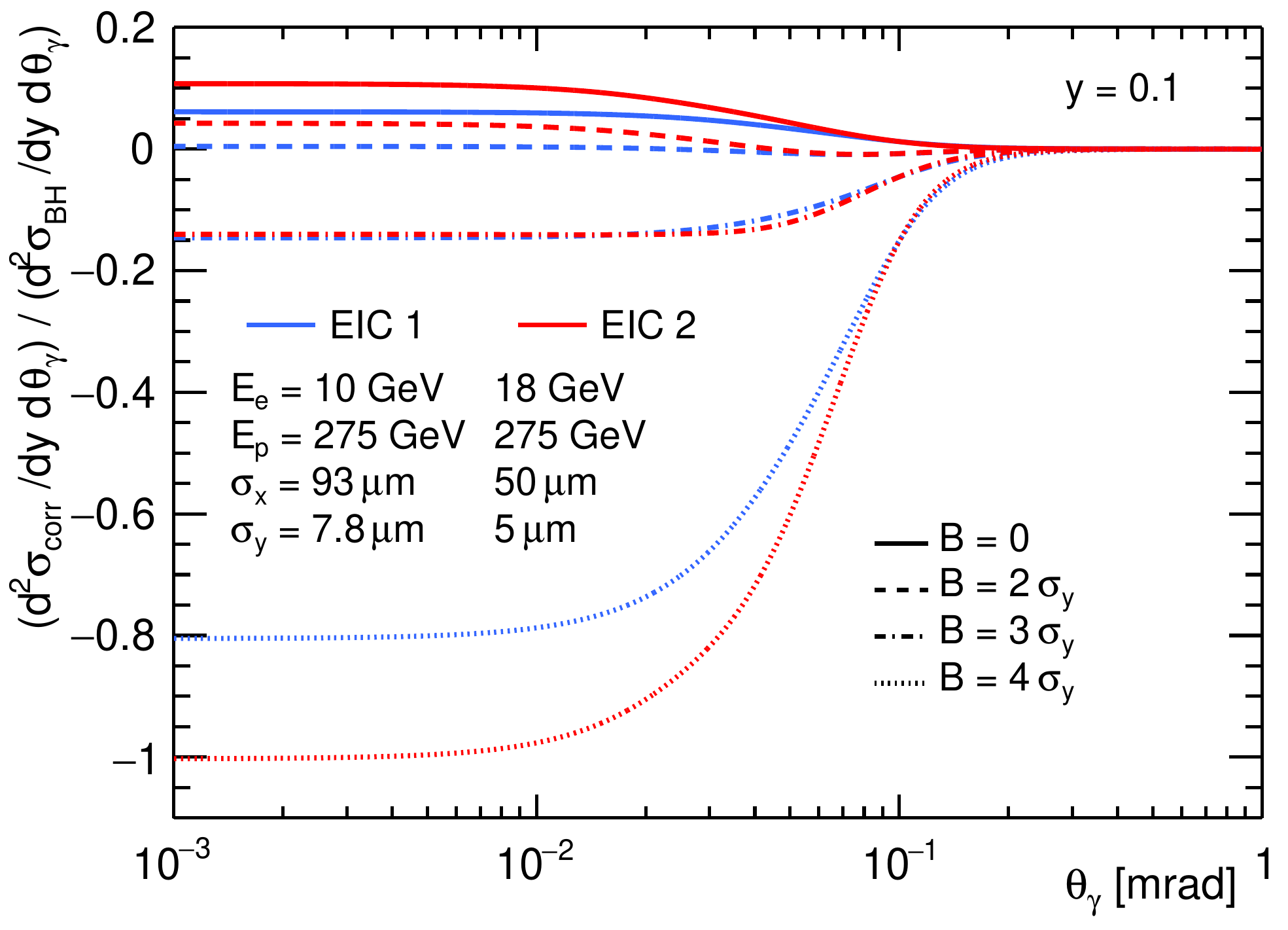}
\caption{\label{fig:fig_7} Relative corrections to the standard Bethe-Heitler cross-sections due to both the beam-size effect and vertical beam displacements. The ratios \dscorrdsBHtheta are shown as a function of the polar angle $\theta_\gamma$ of the photon emission, for four vertical beam displacements $B = 0, 2\sigma_y, 3\sigma_y$ and $4\sigma_y$, for the two sets of EIC parameters - EIC 1 and EIC 2 - the corresponding beam energies and Gaussian lateral beam-sizes at the interaction point are listed. The vertical beam displacements $B = 2, 3$ and $4\sigma_y$ correspond, respectively, to $36, 11$ and $2\%$ of the nominal luminosity for $B = 0$. Plots above are made for (left) $y = 0.001$ and (right) $y = 0.1$, where \ydef.}
\end{figure*}

\begin{figure*}[t]
\includegraphics[width=0.48\textwidth]{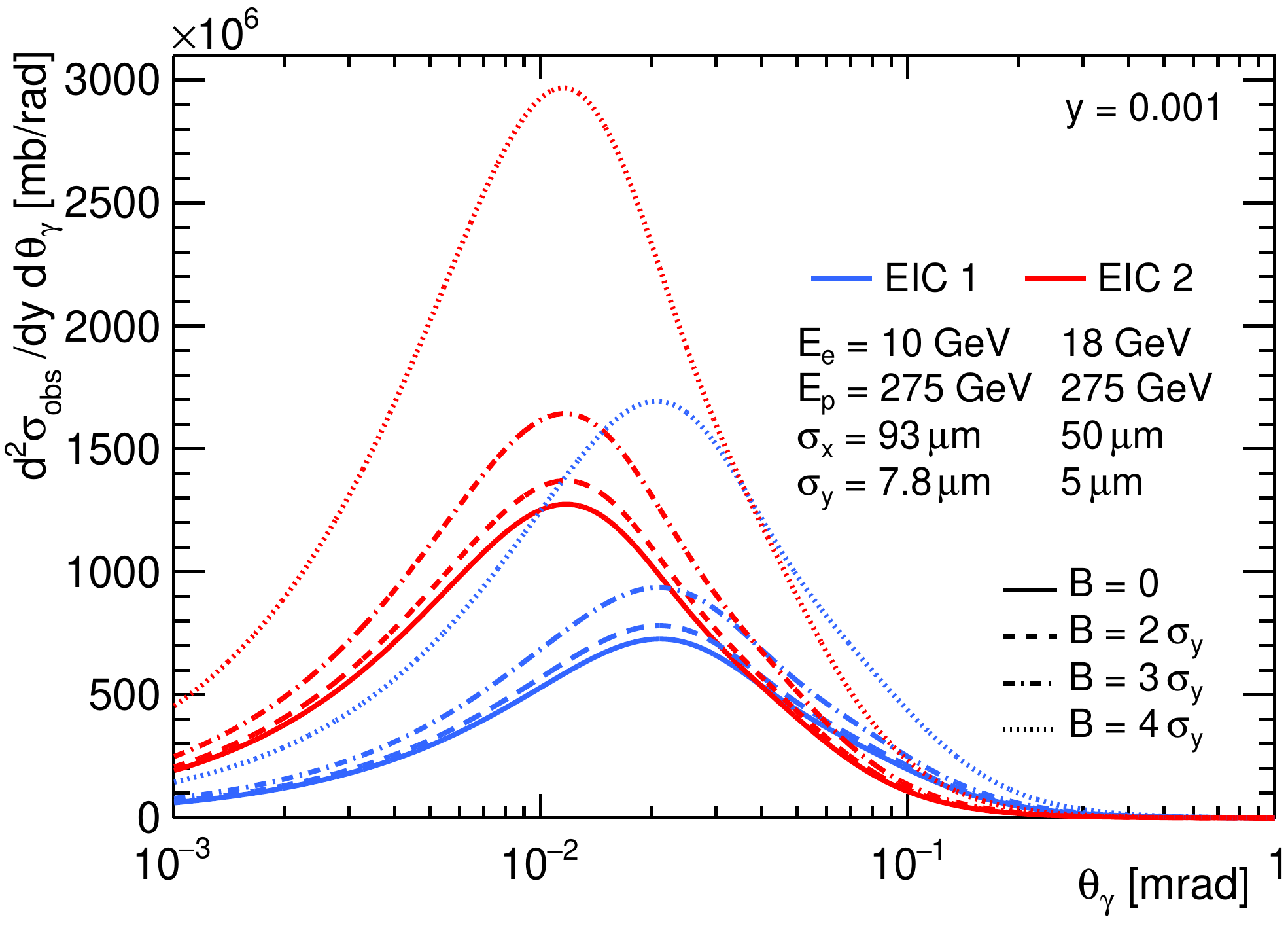}
\includegraphics[width=0.48\textwidth]{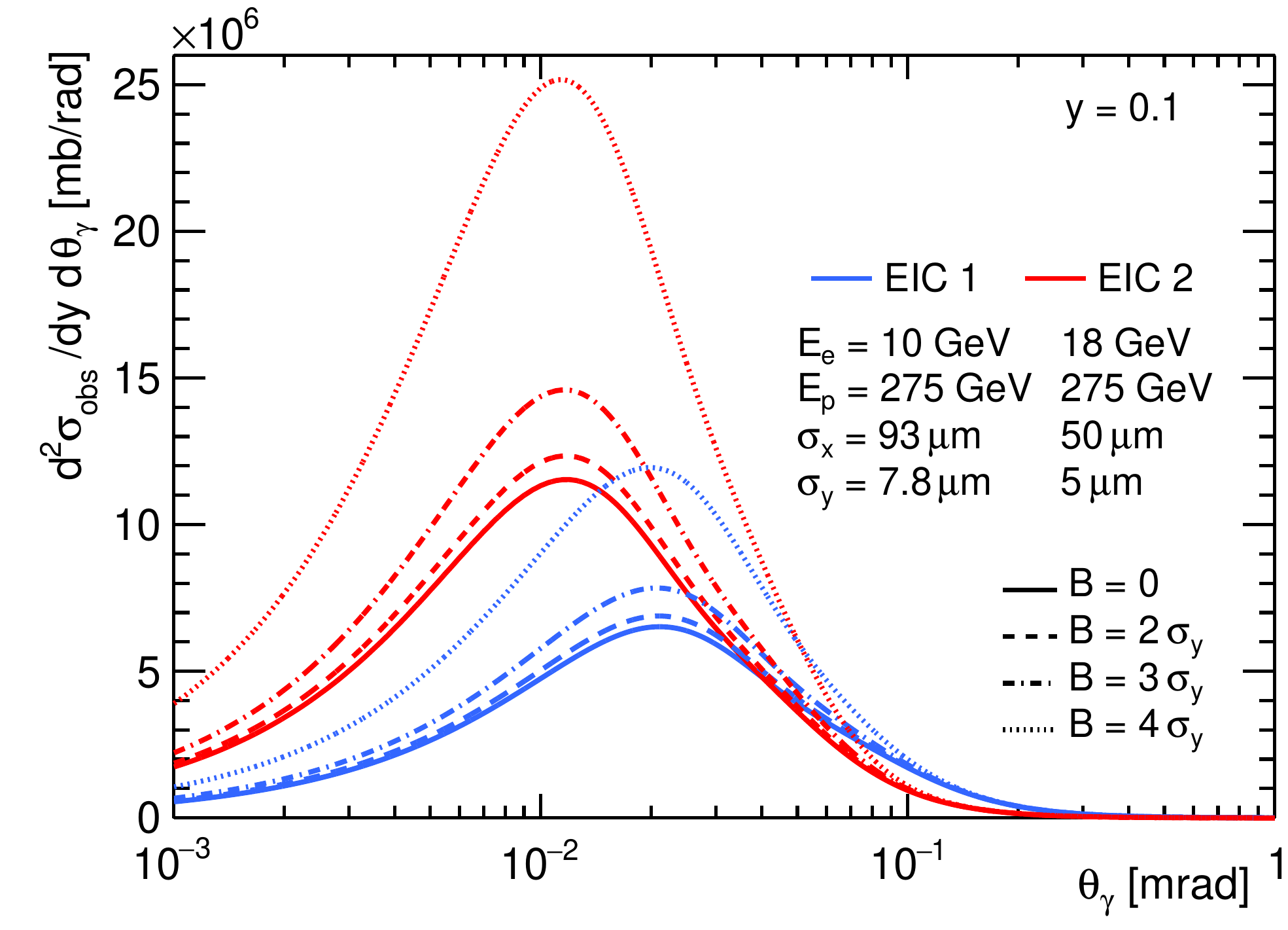}
\caption{\label{fig:fig_8} The predicted photon angular distributions of $ep$ bremsstrahlung at the EIC for several vertical beam displacements. The observed differential cross-sections $d^2\sigma_\mathrm{obs}/dy\,d\theta_\gamma$ are shown as a function of the polar angle $\theta_\gamma$ of the photon emission, for four vertical beam displacements $B = 0, 2\sigma_y, 3\sigma_y$ and $4\sigma_y$, for the two sets of EIC parameters - EIC 1 and EIC 2 - the corresponding beam energies and Gaussian lateral beam-sizes at the interaction point are listed. The vertical beam displacements $B = 2, 3$ and $4\sigma_y$ correspond, respectively, to $36, 11$ and $2\%$ of the nominal luminosity for $B = 0$. Plots above are made for (left) $y = 0.001$ and (right) $y = 0.1$, where \ydef.}
\end{figure*}

%
The studies performed in this paper base on the framework developed in Ref.~\cite{Kotkin:1992bj}. For convenience we provide the relevant formulas here (we keep the formulas' numbering from the original paper). 
The Bethe-Heitler cross section for unpolarized electron and proton beams is given by:\\[-15pt]
\begin{equation}
    \frac{d\sigma_\mathrm{BH}}{dy} = \frac{16\alpha^3}{3m_e^2}\frac{1}{y}\left(1-y+\frac{3}{4}y^2\right)L
    \tag{6.5}
    \label{eq:eq_dsBH}
\end{equation}
where $\alpha$ is the fine structure constant, $m_e$ is the electron mass, \ydef, and
\begin{equation*}
L=\ln{\frac{4E_eE_p(1-y)}{m_em_py}}-\frac{1}{2}
\end{equation*}
The correction to the nominal Bethe-Heitler cross section, where it is assumed that $d\sigma_\mathrm{obs}/dy = d\sigma_\mathrm{BH}/dy - d\sigma_\mathrm{corr}/dy$, is obtained from:
\begin{equation*}
    \frac{d\sigma_\mathrm{corr}}{dy} = 2\,\frac{\alpha^3}{m_e^2}\frac{1}{y}\int_0^\infty{\!F(y,z)\,G(\omega)\,\frac{dz}{(1+z)^2}}
    \tag{6.7$^\star$}
    \label{eq:eq_dscorr_new}
\end{equation*}
which is twice as big as the correction in Eq. (6.7) in Ref.~\cite{Kotkin:1992bj} due to summing over photon helicities, and where
\begin{equation}
    F(y,z) = 2-y-4\frac{(1-y)z}{(1+z)^2}-y(1-y) \tag{6.3}
\end{equation}
and $G$ is the function which describes the number of missing equivalent photons of energy $\omega$:
\begin{equation}
    G(\omega) = \int_0^\infty{\!K_1^2(\rho_\perp/\rho_m)\,g(\vec{\rho}_\perp)\,\frac{d\vec{\rho}_\perp}{\pi\rho_m}} \tag{5.18}
\end{equation}
where
\begin{equation*}
    \omega = \frac{m_e^2 y (1+z)}{4E_e(1-y)} \;\;\;\;\;\mbox{and}\;\;\;\;\; \rho_m = E_p/(m_p \omega)
\end{equation*}
 The predictions obtained in Ref.~\cite{Kotkin:1992bj} assumed head-on beam collisions and Gaussian distributions of the beam particle coordinates in the transverse plane, $n(\vec{r}_\perp)$. We extended this framework by allowing relative, lateral beam displacements. Above we discussed the vertical displacements and here we introduce a horizontal beam displacement $B$ of one of the colliding beams:
%
$$
n(\vec{r}_\perp) = \frac{N}{2\pi \sigma_x\sigma_y}\exp{\left(-\frac{(x-B)^2}{2\sigma_x^2}-\frac{y^2}{2\sigma_y^2}\right)}
$$
%
\noindent
where $\vec{r}_\perp =(x, y, 0)$, $N$ is the number of particles in a beam and \sx, \sy are the standard deviations in the horizontal and vertical planes, respectively. This resulted in the replacement of the original Eq. (5.32) by the following formula:
\begin{widetext}
\begin{equation}
G(\omega) = 2\int_{0}^{\infty}{\frac{\rho_\perp}{\rho_m} K_1^2\left(\frac{\rho_\perp}{\rho_m}\right)\left[1-\frac{e^{-v_+}}{\pi}\int_0^\pi{e^{v_-\cos{\varphi}}\cosh{\left(t_x\cos{\left(\frac{\varphi}{2}\right)}\right)d\varphi}}\right]}\frac{d\rho_\perp}{\rho_m}
\label{eq:eq_4}
\end{equation}
\end{widetext}
\noindent
where 
$t_x =\rho_\perp B/a_x^2$ and $v_\pm = \rho_\perp^2 (1 \pm a_y^2/a_x^2)/(4a_y^2)$, where $a_x^2 = \sigma_{x1}^2 + \sigma_{x2}^2$ and $a_y^2 = \sigma_{y1}^2 + \sigma_{y2}^2$ and $\sigma_{x1}$, $\sigma_{y1}$ and $\sigma_{x2}$, $\sigma_{y2}$ are, respectively, the horizontal and vertical beam sizes at the interaction point (that is, at $z = 0$) for two colliding beams; $K_1$ is the modified Bessel function of the second (third) kind.\\
\indent
The plots in Fig.~\ref{fig:fig_6}~(left) below are effectively obtained by making cross-sections of the two-dimensional distributions shown in Fig.~\ref{fig:fig_5} at fixed $B$ - shapes of the obtained curves are driven by the beam displacement effect at high $y$, and by the beam-size effect at low $y$. In Fig.~\ref{fig:fig_6}~(right) the corresponding plots 
are shown for the horizontal displacement $B$ at the EIC, according to Eq.~\eqref{eq:eq_4} above. Finally, for completeness, in Figs.~\ref{fig:fig_7} and \ref{fig:fig_8} we show modifications of the photon angular distributions due to the vertical beam displacements and lateral beam-sizes.\\
\indent
All plots have been produced using the \texttt{ROOT} 
analysis framework \cite{Brun:1997pa}. Numerical integrations were performed using GNU Scientific Library 
\cite{GSL} interfaced to \texttt{ROOT}. The library reimplements the algorithms used in \texttt{QUADPACK} which are described in Ref.~\cite{QUADPACK}.\\
%
%
\end{document}